# Ion Type Dependence of DNA Electronic Excitation in Water under Proton, α-particle, and Carbon Ion Irradiation: A First-Principles Simulation Study

Christopher Shepard[1] and Yosuke Kanai[1,2]*

[1]Department of Chemistry, University of North Carolina at Chapel Hill, Chapel Hill, North Carolina 27514, USA

[2]Department of Physics and Astronomy, University of North Carolina at Chapel Hill, Chapel Hill, North Carolina 27514, USA

*Supporting Information Placeholder*

**ABSTRACT:** Understanding how the electronic excitation of DNA changes in response to different high-energy particles is central to advancing ion beam cancer therapy. While protons have been the predominant ion of choice in beam cancer therapy, heavier ions, particularly carbon ions, have drawn significant attention over the past decade. Carbon ions are expected to transfer larger amounts of energy accordingly to the linear response theory. However, the molecular-level details of the electronic excitation under heavier ion irradiation remain unknown. In this work, we use real-time time-dependent density functional theory simulations to examine the quantum-mechanical details of DNA electronic excitations in water under proton, α-particle, and carbon ion irradiation. Our results show that the energy transfer does indeed increase for the heavier ions, while the excitation remains highly conformal. However, the increase in the energy transfer rate, measured by electronic stopping power, does not match the prediction by the linear response model, even when accounting for the velocity-dependence of the irradiating ion's charge. The simulations also reveal that while the number of holes generated on DNA increases for heavier ions, the increase is only partially responsible for the larger stopping power. Larger numbers of highly energetic holes formed from the heavier ions also contribute significantly to the increased electronic stopping power.

## 1. Introduction:

The electronic excitation response of DNA to high energy ions is the foundation of modern ion beam cancer therapy. In beam cancer therapy, irradiating ions with large (MeV) kinetic energies penetrate through the body.[1] As the ions slow down and transfer their kinetic energy by exciting electrons, a sharp increase in the energy transfer is observed at the very end of the ion's penetrating range. This distinctive peak, known as the Bragg peak, in the energy deposition profile of ions derives from the electronic stopping power, or the velocity-dependent energy transfer rate from the ion to the target material, and is what initially drew interest for radiation oncology.[2] The highly localized energy deposition profile is unlike that of traditional photon-based radiation, in which significant amounts of energy are deposited at the beginning of the photon's penetration. This characteristic energy deposition profile allows high-energy ion radiation to be tuned to focus directly on the tumor, with minimal damage to surrounding healthy tissues.[3] Over the last few decades ion beam cancer therapy has emerged as a viable alternative to conventional x-ray oncology treatments.[4-5] In particular, small ions have proven to be effective for treating tumors near sensitive areas of the body.[6-9] Studies have indicated that these ions generate clustered localized lesions, including double-strand breaks (DSB) of DNA, that are likely to lead to cell death.[10-12] More recently, carbon ions have drawn increased attention due to the larger energy transfer expected at the Bragg peak, while still maintaining a highly conformal dose deposition.[13-16]

However, there are still many unanswered questions in ion beam therapy. In practice, the given dose is calibrated by scaling the relative biological effectiveness (RBE) from traditional photon therapy by the absorbed dose from the ion beam in water.[17-18] This assumes DNA damage under ion radiation is the same as that under photon radiation, an assumption that has previously been called into question.[18-19] Recent studies have shown that the RBE depends on many factors, including particle path, particle type, and the electronic stopping power of the particle, with such complexity that there are inconsistencies in how the biological weighing factor is currently calculated and applied.[20-22] Further investigation is necessary to improve the scientific understanding of the DNA damage process and create a consistent model for ion beam calibration.[21, 23-24]

Over the past ten years, many studies have begun to take aim at deciphering the complex DNA damage process, and improve these models.[25-26] Sophisticated multi-scale models, such as Geant4-DNA and KURBUC, have provided significant insight into the physical and chemical interactions in ion beam radiotherapy.[27-28] However, many of the molecular-level details of the damage process are still unknown. The high cost of ion beam center construction and operation[29], coupled with the ultrafast nature of the excitations, has made experimental investigation difficult.[30] Additionally, available experimental results are still relatively limited, especially for ions other than protons.[21, 31] First-principles theoretical modeling plays an essential role in filling the knowledge gaps of how electronic excitation leads to DNA damage on the molecular level.[32]

With the increase in peta and exa-scale supercomputers[33] along with highly efficient and massively parallel first-principles electronic structure codes[34-36], direct simulation of the electronic excitation dynamics under ion radiation is now possible.[32] Specifically, real-time time-dependent functional theory (RT-TDDFT) presents an avenue to directly simulate the quantum-dynamics electronic response of complex systems, such as DNA in water, under ion radiation.[37-38] In our recent work, this approach was demonstrated in studying the electronic excitation response of DNA in water under proton irradiation.[39] In this work, we examine how the electronic excitation dynamics of DNA change under $\alpha$-particle and carbon ion irradiation in water, as the medical community hopes to obtain better outcomes with these higher-Z ions in ion beam cancer therapy.

## 2. Computational Method:

In this work, we use our RT-TDDFT implementation within the Qb@ll branch of the Qbox code[40-41], based on the planewave pseudopotential formalism.[42] The solvated DNA structure is identical to that used in our previous works,[39, 43] a strand of B-DNA containing 10 base pairs (sequence CGCGCTTAAG) and comprising one full turn of the double helix.[44] The simulation cell, with periodic boundary conditions, contains the full turn in the z-direction, and the simulation cell is made commensurate with the periodicity of the macromolecule. The cubic simulation cell, with dimensions of 34.43 Å, is large enough to prevent errors stemming from the periodic images in the x and y directions.[39] The DNA strand is solvated with 1119 explicit water molecules. Further details on the simulation cell and equilibration can be found in the "Computational Methods" section of the Supporting Information, as well as in reference 39. All simulations include 3991 atoms (634 for DNA and 3357 for water) and 11,172 electrons, requiring the use of the highly scalable and massively parallel aforementioned Qb@ll code[36] and peta-scale computing resources to perform the calculations. The RT-TDDFT simulations necessitated the use of up to 262,144 Intel KNL 7230 processor cores on the Theta supercomputer at the Argonne leadership Computing Facility. All atoms are represented by Hamann-Schluter-Chiang-Vanderbilt (HSCV) norm-conserving pseudopotentials.[45-46] The Perdew-Burke-Ernzerhof (PBE)[47] generalized gradient approximation (GGA) functional was used for the exchange-correlation approximation, along with a planewave cutoff energy of 50 Rydberg.

Both $\alpha$-particle and carbon "projectile" ions are treated using HSCV pseudopotentials, just as the irradiating proton was in our previous study.[39] In the case of the proton projectile, the initial state is fully ionized, and the proton was shown to reach a velocity-dependent effective charge state in water prior to interaction with DNA (see Figure S1 in the Supporting Information). However, heavier ions such $\alpha$-particles and carbon ions often do not reach a steady charge state in water prior to interaction with DNA in the simulation cell, depending on the ion's velocity. Without a correct steady-state charge, fictitious charge transfer between the projectile ion and DNA can result in an unrealistic situation.[48-49] In order to overcome this difficulty, the $\alpha$-particle and carbon ions are given a velocity-dependent initial charge. To set the initial charge of the projectile ions, the conventional time-dependent Kohn-Sham (TD-KS) states were expressed in the Maximally Localized Wannier Function (MLWF) gauge[50], including that of the projectile ion. MLWFs are spatially localized on chemical moieties, offering a chemically intuitive picture of the system along with a localized orbital for the projectile ion, without changing the underlying electron dynamics. At the beginning of each simulation, we identify the MLWF spatially localized on the projectile ion. By changing the occupation number for the projectile-localized MLWF, a specified velocity-dependent electronic charge can be assigned to the projectile ion. We determine this initial charge from a separate series of RT-TDDFT simulations that determine the velocity-dependent, mean steady-state charge for the specific ion in liquid water (see Figures S2-S3 in the Supporting Information for details). To ensure the MLWF associated with the projectile is moved with the projectile ion, the same velocity is applied to the MLWF through an impulsive electric field. The impulsive electric field is applied as the initial phase of the projectile's Wannier function, so that only the Wannier function associated with the projectile is affected. The MLWFs are propagated as time-dependent maximally localized Wannier functions (TD-MLWFs)[50-51]. All simulations use a 2.0 attosecond time step and the enforced time reversal symmetry propagator[52] for integrating the electronic states in time. The positions of all DNA and water atoms are held constant, as the time scales for these simulations (0.27-3.38 fs, depending on projectile ion velocity) is too short for any notable nuclear motion.[53] The projectile ion is moved at the constant velocity of interest for obtaining the velocity-dependent electronic stopping power curve.[32, 54] The electronic stopping power, or the energy transfer rate from projectile ion to target, is a crucial property for many applications, including beam cancer therapy.[55-57] As the projectile ion travels through the simulation cell, the electron density changes in response to the time-dependent potential generated by the projectile ion until the ion reaches the end of its trajectory and the simulation is stopped. By moving the projectile at a constant velocity, while all other atoms are held in place, the total energy of the non-equilibrium simulation is not conserved, as work is done throughout by the projectile ion.[32, 42, 54] Therefore, changes in the total energy of the system can be used to calculate the electronic stopping power as a function of projectile ion velocity (see Figures S4-S5 for details).[36]

## 3. Results and Discussion:
### 3.1 Electronic Stopping Power

Figure 1 shows the DNA strand fully solvated in water, with the simulation cell outlined in black. Each panel shows changes in the electron density in response to an irradiating (a, d) proton, (b, e) $\alpha$-particle, and (c, f) carbon ion. For direct comparison to our previous work on proton irradiation[39, 43], we consider two paths for the projectile ions; the Base path directly through the center of the DNA strand (shown in orange in Figure 2(a) and Figure S6 with the Supporting Information) and the Side path along the side of the sugar-phosphate side chain (shown in purple in Figure 2(a) and Figure S6). In order to compare $\alpha$-particle and carbon ion irradiation to the case of proton irradiation, we considered the same six velocities from our previous work on proton irradiation in DNA[39]. For the $\alpha$-particle, we performed additional simulations with the velocities at the stopping power maximum (Bragg peak, v = 2.27 a.u.) in dry DNA[43] and at the Bragg peak (v = 2.44 a.u.) in liquid water.[58] For the carbon ion, we did not consider the velocity of 6.0 a.u. because the carbon ion at this velocity is close to fully ionized and additional treatment, including a much larger (and not currently possible) plane-wave cutoff energy, would be needed to properly model the core electrons.[59-61] We simulated the Bragg peak velocity for the carbon ion in liquid water[62] (v = 3.17 a.u.) and the Bragg peak for the carbon ion in dry DNA (v = 2.44 a.u.), (see Figure S7 in Supporting Information for details). The velocity of 3.17 a.u. was also simulated for the proton and $\alpha$-particle cases as an additional point of comparison. Comparison of the electronic stopping power curves for the Base path (Figure 2(b) solid lines) and the Side path (Figure 2(c) solid lines) show that stopping power magnitude for all ions is notably larger for the Side path. For the $\alpha$-particle, the stopping power along the Side path shows a magnitude that is more than three times greater at the Bragg peak and at least twice as large at all velocities studied here, when compared to the Base path (see Figure S8 for direct comparison). The difference between the two paths is smallest for the higher (4.00 and 6.00 a.u.) velocities. For the carbon ion, the stopping power magnitude is twice as large at the peak and at least 1.5 times greater throughout the curve for the Side path, when compared to the Base path (see Figure S8 for direct comparison). For both the Base and Side paths, the Bragg peak velocities depend on irradiating ion type; the $\alpha$-particle and carbon ions show the Bragg peak at higher velocities than the irradiating proton. This trend is consistent with previous observations for $\alpha$-particle irradiation on dry DNA.[43]

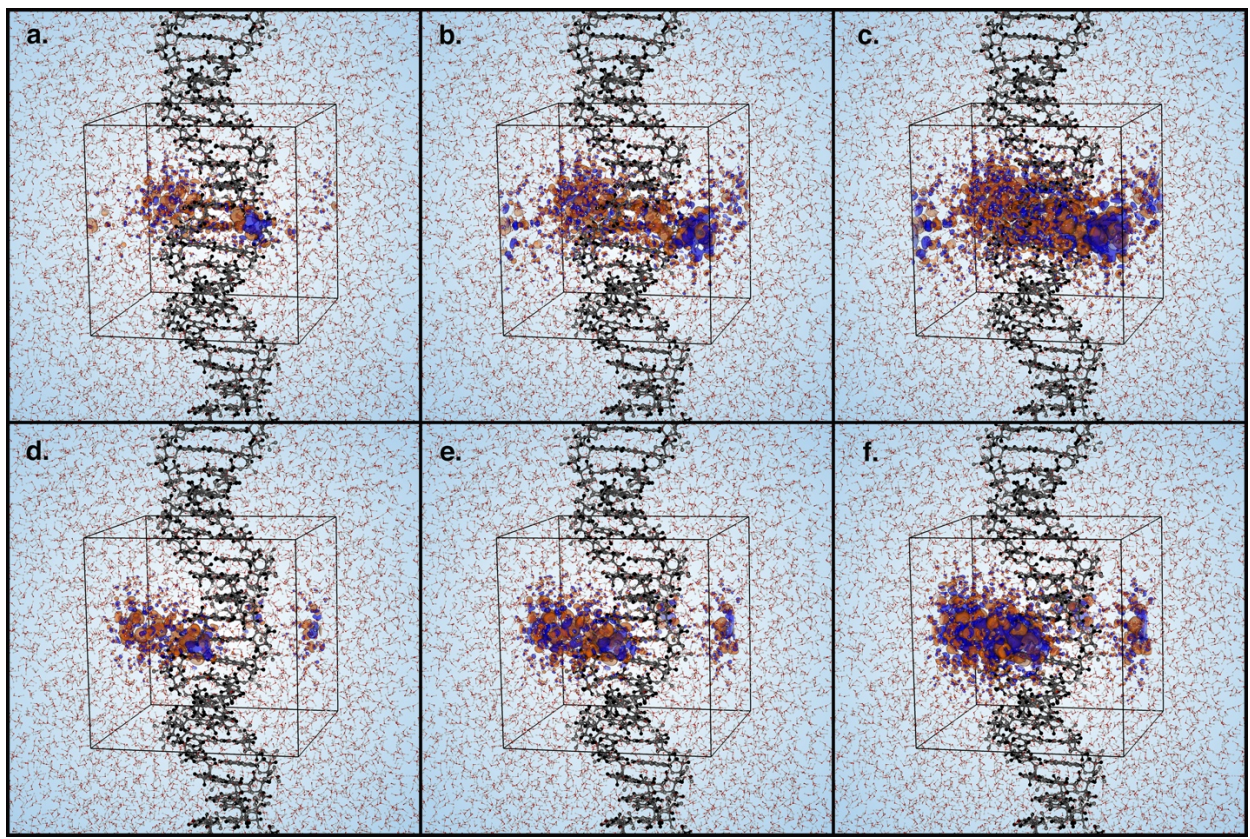

**Figure 1.** RT-TDDFT simulation snapshots for a (a) proton moving at 1.64 a.u. velocity, (b) $\alpha$-particle moving at 2.50 a.u. velocity and (c) carbon ion moving at 2.44 a.u. velocity along the Base path, and for a (d) proton moving at 1.64 a.u. velocity, (e) $\alpha$-particle moving at 1.90 a.u. velocity and (f) carbon ion moving at 1.90 velocity, along the Side path. The simulation cell, outlined in black, is shown with periodic boundary conditions for solvated DNA. Blue (orange) isosurfaces represent increases (decreases) in electron density relative to the ground-state electron density, at the end of each projectile path. For comparison, the same isosurface values are used in all cases. Velocities correspond to points closest to the irradiating ion's stopping power maxima.

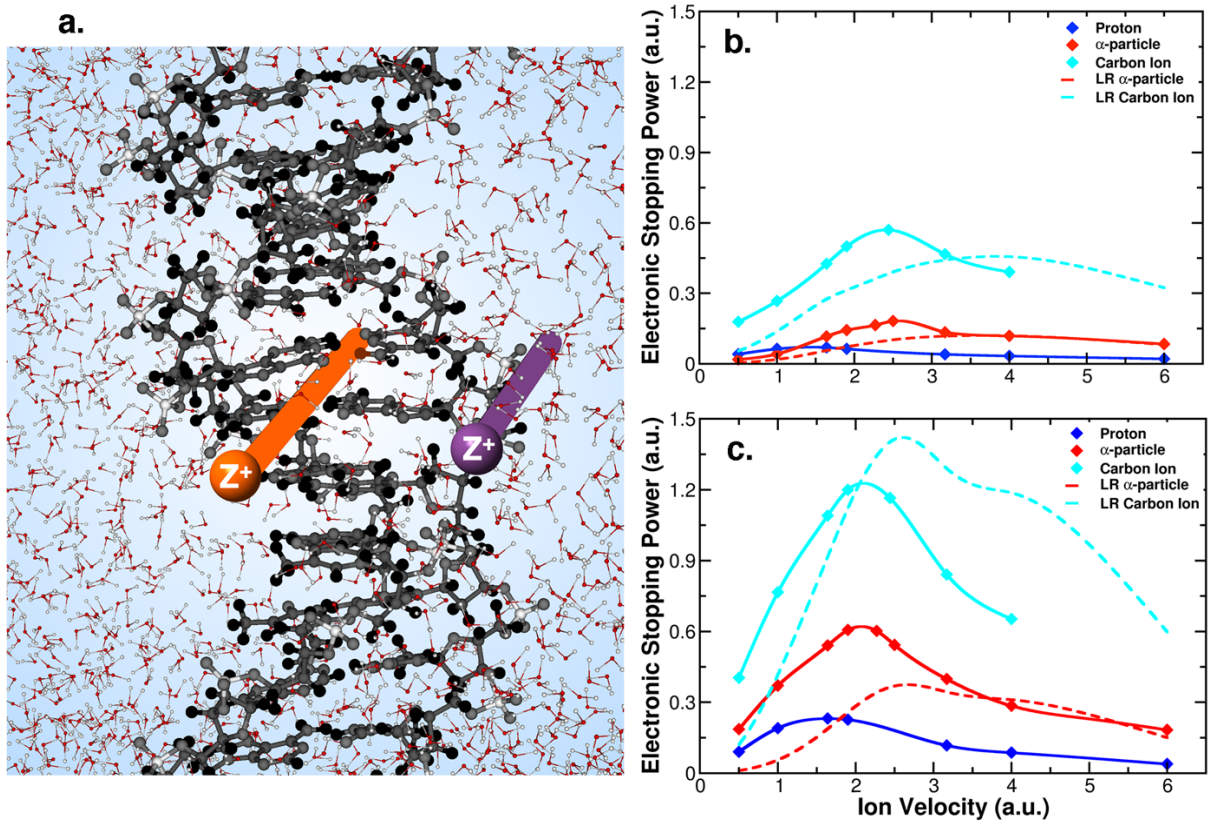

**Figure 2.** (a) Solvated DNA structure, with the Base path denoted by the orange line and the Side path denoted by the purple line. Electronic stopping power for the (b) Base and (c) Side paths, calculated as the average instantaneous stopping power over the DNA-interaction region (see Figures S4-S5 for details), for an irradiating proton (blue), $\alpha$-particle (red) and carbon ion (cyan). Linear response scaled electronic stopping power curves (dashed lines and denoted LR) are determined by scaling the proton stopping power curves by a factor of $Z(v)^2$, where $Z(v)$ is the effective charge of the ion for a given velocity in water (see Figure S3 for details).

Linear response theory is widely used for determining stopping power,[63] and it is of interest to examine this approximate description in the light of our first-principles dynamics result. The observed shift in the Bragg peak velocity among different projectile ions is beyond the description based on linear response (LR) theory, which can be expressed as:

$$S(v) = \frac{4\pi Z^2}{v^2} L(v) \qquad (1)$$

where $S(v)$ is the electronic stopping power, $v$ is the projectile ion velocity, and $L(v)$ is the velocity-dependent term known as the stopping logarithm.[64-65] There exist a number of different expressions for the stopping logarithm[66-67], which incorporates information of the target material only. Generally, in LR theory, the charge of the projectile ion is treated as a velocity-independent quantity and ions are assumed to be fully ionized.[58, 68] This assumption leads to well-known issues such as incorrectly predicting the same Bragg peak position for all ions.[58, 69-70]

Instead, the ion charge can be treated as a function of ion velocity, $Z(v)$, being equal to the mean charge for each ion in liquid water for this study. This quantity is equivalent to the charge given to each ion at the start of the RT-TDDFT simulations (see Figure S3 in Supporting Information). In the LR description, the electronic stopping power depends only on $Z(v)$ for the same target matter. Figures 2(b) and 2(c) show the resulting LR-scaled stopping power curves for the $\alpha$-particle and carbon ion, using the stopping power for the proton as the stopping logarithm (dashed lines) and denoted as the "LR"-scaled model within the figure. For $\alpha$-particle velocities in the high-velocity regime (at and above 4.00 a.u.), where the effective charge of the ion is or is close to fully ionized, the LR-scaled model matches well with the RT-TDDFT results. However, closer to the Bragg peak the LR-scaled model and the RT-TDDFT results begin to diverge, with the LR-scaled model underestimating the stopping power magnitude by as much as 40% and overestimating the peak position. We also note that at low velocities (0.5 and 1.0 a.u.) the LR model for the Side path differs significantly from the RT-TDDFT result. For the carbon ion, the LR-scaled model differs from the RT-TDDFT results significantly at all velocities, overestimating the stopping power at higher velocities and underestimating the stopping power at lower velocities. The LR-scaled model for the carbon ion also fails to reproduce the Bragg peak position of the RT-TDDFT result. Even when accounting for the velocity dependence of the ion charge $(Z(v)^2)$ term, scaling the electronic stopping power of the proton using the LR theory does not correctly predict the energy transfer rate for heavier ions. This result is particularly important for applications of irradiation using heavy atoms, such as carbon ions, for ion beam therapy.[13]

### 3.2 Molecular Decomposition of Electronic Excitations

We discuss here the underlying electronic excitation responsible for the observed differences in the electronic stopping power among the different ion radiation types, as it may be particularly useful for deciphering DNA damage at the molecular level.[39] As discussed above, the RT-TDDFT simulations were performed using the MLWF gauge[50-51] such that the electronic

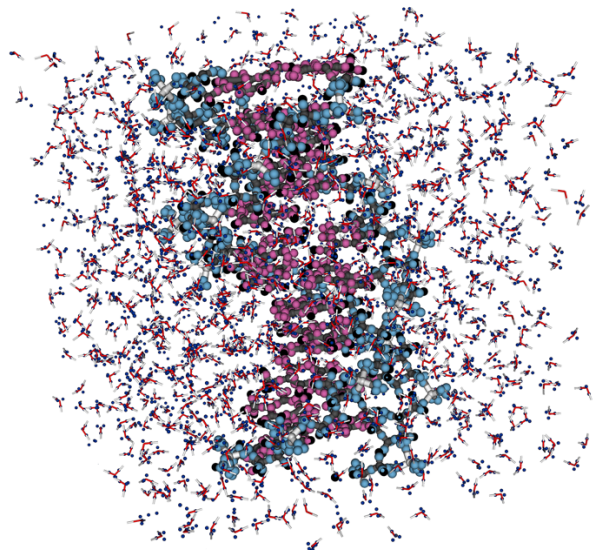

**Figure 3.** Solvated DNA structure, with ground state MLWFs shown as dark blue spheres (water), magenta spheres (nucleobases) and light blue spheres (sugar-phosphate side chains).

excitation response can be decomposed in terms of the molecular constituents. As shown in Figure 3, the geometric centers of the TD-MLWFs (often called Wannier centers) are spatially localized on different chemical moieties. The electronic response of the solvating water molecules can be separated out from that of the DNA, as the TD-MLWFs can be grouped into different subgroups. We further divide the DNA-localized TD-MLWFs into two chemical subgroups, nucleobases (shown in magenta in Figure 3) and sugar-phosphate side chains (shown in light blue in Figure 3) to analyze how the induced excitations effect different subgroups. Previous work on dry DNA showed further decomposition did not yield additional physical insights.[43] To compare the excitations for different ion radiation types, we analyze the electronic excitation response in terms of changes to the positions and spatial spreads of the TD-MLWFs.[39] Figure 4 shows the comparison of Wannier center displacements and spread changes for the projectile proton (a, d), $\alpha$-particle (b, e) and carbon ion (c, f) at the 1.90 a.u. velocity. The 1.90 a.u. velocity is used here for comparison as it is the closest data point to the Bragg peak for both the $\alpha$-particle and carbon ion along the Side path, which shows a significantly larger stopping power than the Base path, as discussed above. Spatial decomposition (Fig. 4) shows that 64-67% of the Wannier center displacements and more than 95% of the spread changes occur within 5.0

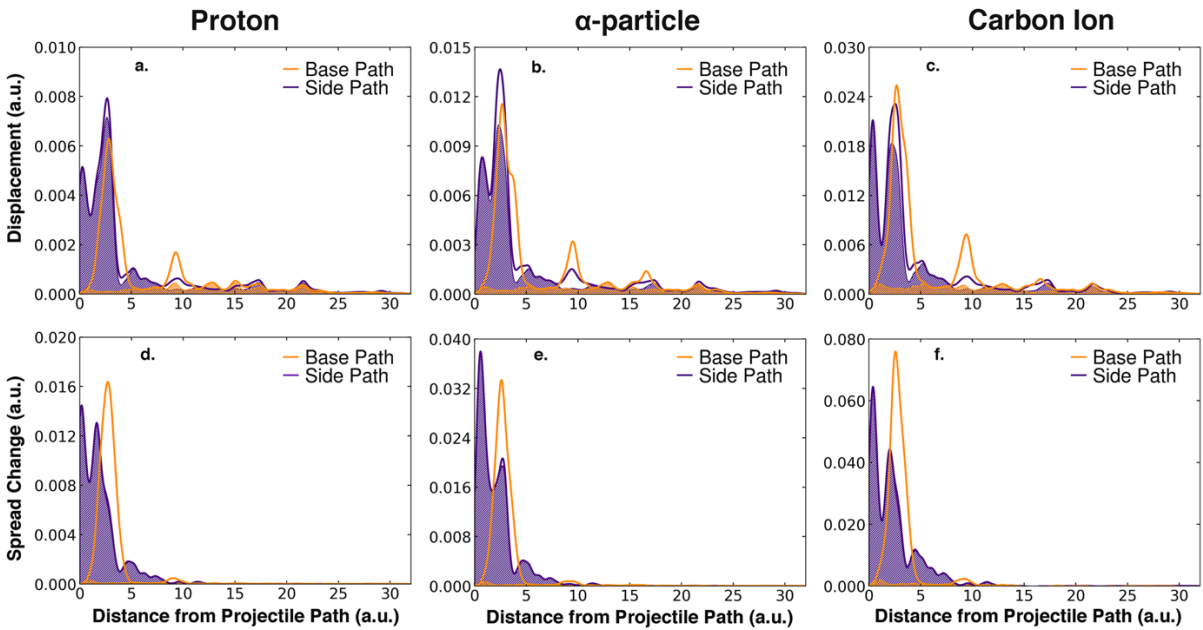

**Figure 4.** Displacement of the DNA TD-MLWF centers in response to an irradiating (a) proton, (b) $\alpha$-particle and (c) carbon ion at 1.90 a.u. velocity. Spread changes of the DNA TD-MLWFs in response to an irradiating (d) proton, (e) $\alpha$-particle and (f) carbon ion at 1.90 a.u. velocity. Hatched regions correspond to contributions from the DNA phosphate side chain.

Bohr of the projectile ion path for all projectile ions along the Base path (orange in Figure 4). Similarly, for the Side path (purple in Figure 4), 70-75% of the Wannier center displacements and more than 88% of the spread changes occur within 5 Bohr of the projectile ion path. While the heavier ions yield larger displacements and spread changes, the electronic excitation is highly localized in the immediate vicinity of the projectile ion path[71], regardless of the irradiating ion type. Figure 4 also shows the changes in terms of the two DNA subgroups, with the hatched areas corresponding to the electronic response from the phosphate side chain. For the Side path, more than 75% of the displacements and almost all (>98%) of the spread changes are from the phosphate side chain for all projectile ions. On the other hand, for the Base path, very minimal changes are observed on the phosphate side chain, as over 75% of the displacements and more than 94% of the spread changes are on the nucleobases. This highly localized nature of the electronic excitations is observed for all velocities above and below the Bragg peak, and for all ion radiation types as well (see Figures S9-S10 in the Supporting Information).

### 3.3 Hole Generation within DNA

Beyond the conformal nature of the excitations, the TD-MLWFs also allow for analysis of the holes generated during ion irradiation. While the hole population (i.e., number density) is often assumed to be directly proportional to the stopping power in literature[72], previous work has shown that this is not the case[39, 43, 71]. At the same time, increased number of holes generated on DNA is thought to be a key factor for the increased damage and the rise in double strand breaks under irradiating ions heavier than protons.[16, 21] To quantify the DNA hole population, we project the DNA-localized TD-MLWFs onto the equilibrium Kohn-Sham (KS) eigenfunctions. The energy-dependent DNA hole population generated by the irradiating ion can be calculated as:

$$HP(t,\varepsilon) = \sum_j^{Nocc} \sum_n \left(2 - f_n |\langle \psi_j | w_n(t) \rangle|^2 \right) \delta(\varepsilon - \varepsilon_j)$$

(2)

where $f_n$ is the occupation of the DNA-localized TD-MLWFs, $w_n(\mathbf{r},t)$, and $\psi_j(\mathbf{r})$ is the KS eigenstate in the valence band with the energy $\varepsilon_j$. The hole population was computed at the end of each simulation trajectory, at which point the DNA hole populations were found to reach essentially a constant value (see Figures S11-S13 in the Supporting Information). Figure 5 shows the total DNA hole population (i.e., summation over the entire energy range) as a function of ion velocity. The hole population on DNA is not the main reason for the electronic stopping power difference between the Base and Side paths because they show the similar magnitudes. Rather, the difference derives from the energetics of the generated holes as discussed already for proton irradiation in Ref. 39. Therefore, we examine how the hole population changes as the irradiating ion is changed from proton to $\alpha$-particle and carbon ion, compared to how the stopping power changes. Figure 6 shows the DNA hole population ratios $HP_\alpha(v)/HP_H(v)$ and $HP_C(v)/HP_H(v)$ along with the stopping power ratios, $S_\alpha(v)/S_H(v)$ and $S_C(v)/S_H(v)$, for the Base path and Side path. For the Base path, the increase in stopping power for the $\alpha$-particle directly follows the increase in DNA hole population, compared to the proton case. The ratio of DNA hole populations between the $\alpha$-particle and proton is nearly identical to the ratio of the stopping power, for all velocities studied. For carbon ion irradiation at low velocities, the stopping power again directly follows the increase in DNA hole population. For higher (v > 1.90 a.u.), however, the increase in stopping power for the projectile carbon ion was much larger than the increase in DNA hole population, relative to the proton case. For the Side path which exhibits much larger stopping power (Fig. 2), the increase in stopping power under $\alpha$-particle irradiation was 1.25 to 1.5 times larger than the increase in hole population, relative to the proton case, for all velocities other than 0.50 a.u.. At the velocity of 0.50 a.u., the increase in stopping power was 1.75 times larger than the hole population increase. With carbon ion irradiation, both the stopping power and DNA hole population were larger relative to the proton case, although no clear relationship between the two was observed.

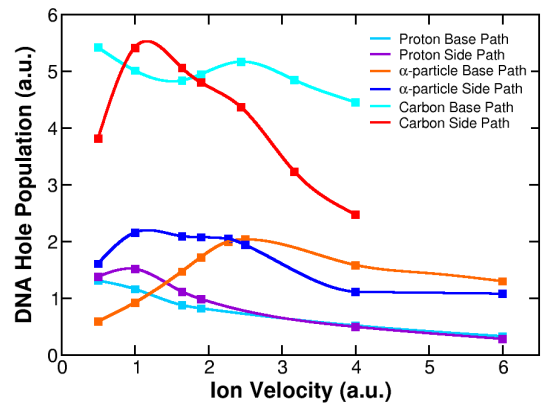

**Figure 5.** Total DNA hole populations, taken at the end of the irradiating ion's path for each irradiating ion, as a function of ion velocity.

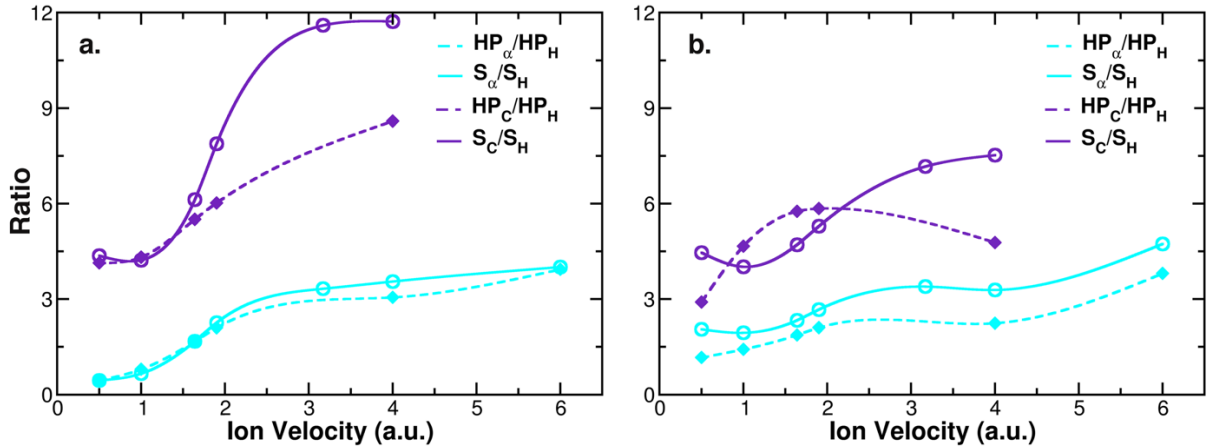

**Figure 6.** Ratio of the calculated electronic stopping power for the $\alpha$-particle (cyan) and carbon ion (purple) to the electronic stopping power of the proton, for the (a) Base path and (b) Side path. Dashed lines correspond to the ratio of the DNA hole population (denoted HP) for the $\alpha$-particle (cyan) and carbon ion (purple) to the DNA hole population for the proton along the (a) Base path and (b) Side path.

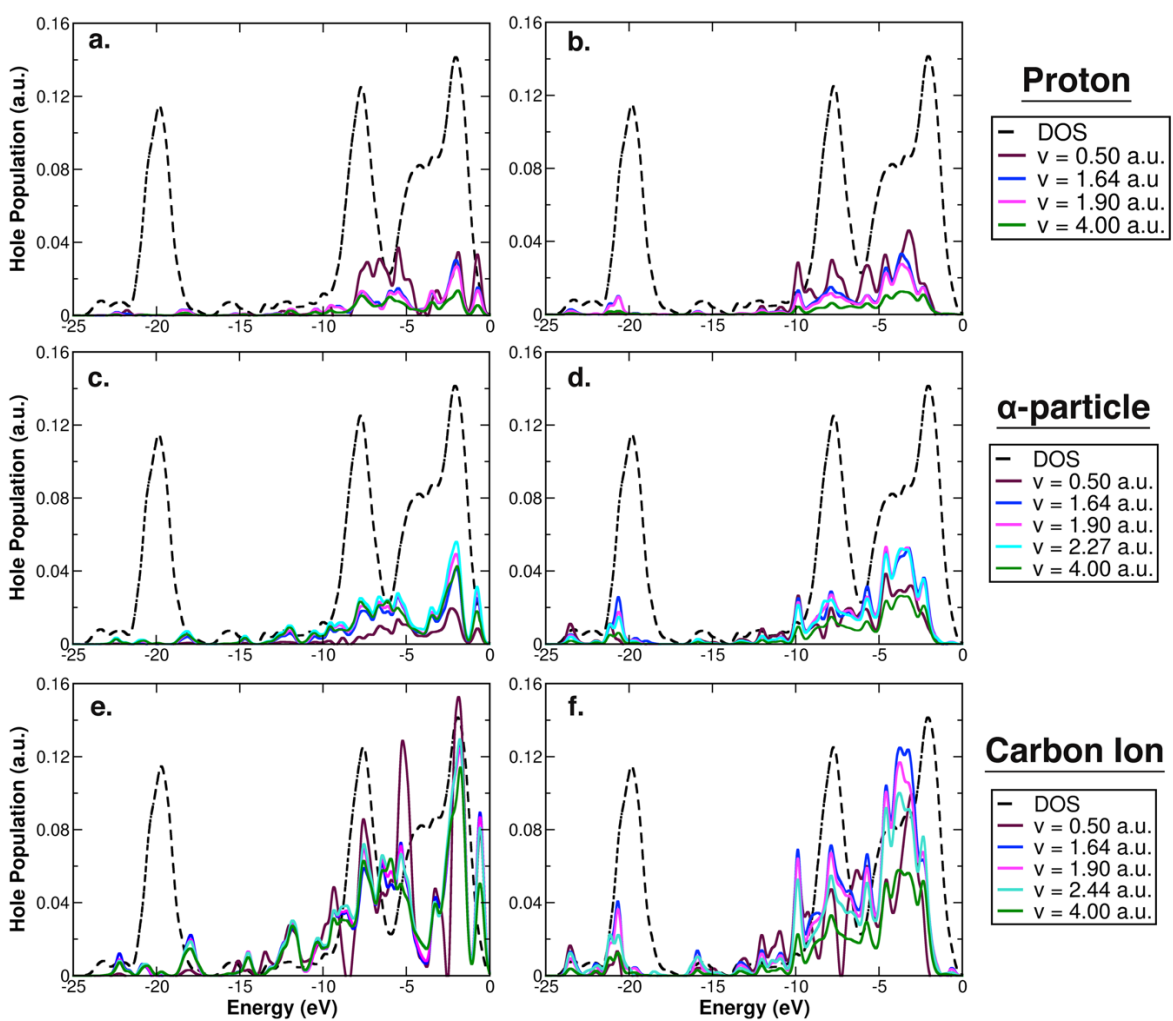

**Figure 7.** DNA hole populations as a function of energy for (a) proton on the Base path, (b) proton on the Side path, (c) $\alpha$-particle on the Base path, (d) $\alpha$-particle on the Side path, (e) carbon ion on the Base path, (f) carbon ion on the Side path. DNA TD-MLWFs are projected onto the energy eigenstates of the system at equilibrium to calculate the energies at which holes are generated in DNA at the end of simulations. Nearly identical energetics were observed at the end of the DNA-interaction region (see Figures S14-S16 for details). For reference, the density of states is shown with a dashed black line. Gaussian broadening of 0.25 eV was used.

The stopping power and the hole population are not directly proportional in all cases, as discussed in our earlier studies, because the energetics of the generated holes factor into the stopping power.[39, 43] Holes formed in deeper energy states have a larger contribution to the stopping power, and it is also conceivable that greater DNA damage might be facilitated by a slow hole relaxation from the deep-lying states that are localized on the phosphate side chains. Our previous work on proton irradiation showed highly-energetic holes were formed only for the Side path.[39] Figure 7 shows the DNA hole population energetics for each irradiating ion type, at a representative set of velocities. The hole energetics are calculated at the end of each simulation trajectory (see Figures S14-S16 in the Supporting Information). The density of states (DOS) is plotted as the dashed black line in Figure 7 for convenience. The highest occupied molecular orbital (HOMO) is set to 0 eV. Among the different ion radiation types, the energies at which holes are formed are similar. As expected, the hole populations are larger for the $\alpha$-particle and carbon ions while other features remain essentially the same.

The hole formation in deeper energy states centered around -20 eV (Figure 7) is of particular importance because the relaxation of the generated holes may be slowed due the large energy separation from the manifold of energy states closer to the valence band maximum. The slow hole relaxation is likely beneficial for oxidative DNA strand damage. For the Side path, both the $\alpha$-particle and carbon ion generate significantly more holes than the proton in this energy range. Figure 7 shows twice as many deep energy holes for the $\alpha$-particle and 4 times as many for the carbon ion at velocities close to the Bragg peak of both heavy ions (1.64-1.90 a.u.), when compared to the proton. For the Base path, the $\alpha$-particle shows very similar hole energetics to the proton case. However, the carbon ion shows some notable differences, with larger amounts of holes generated in the deep-lying states, and those just above -20 eV, for velocities close to and above the Bragg peak. This increase in the deep-energy holes, observed at velocities 1.64 a.u. and above, is the reason for the increase in stopping power not directly attributable to hole population, as seen in Figure 6(a). Just as in the case of proton irradiation[39], significantly more DNA holes are generated in these deep-energy states for the Side path when compared to the Base path for the irradiating $\alpha$-particle and carbon ion, and are responsible for the difference in stopping power between the two paths.

Finally, the hole energetics at velocities far from the Bragg peaks (e.g., 0.50 and 4.00 a.u.) merit some discussion since nonintuitive behavior is observed for the electronic stopping power in the low-velocity range (Figure 2), where significant differences for the stopping power are observed between the different irradiating ions. For proton irradiation, holes are not generated in the deep-energy states at these ion velocities. However, both the $\alpha$-particle and carbon ion generate holes in the deep-energy states for v=4.00 and 0.50 a.u. along the Side path, being responsible for ~5% of the total hole population. For the Base path few holes are formed in the deep-energy states for both $\alpha$-particle and carbon ion irradiation, similar to the response under proton irradiation. Nevertheless, stark differences among the radiation ion types exist at v=0.50 a.u.. As seen in Figure 7 (Left panels), $\alpha$-particle irradiation generates fewer holes than proton irradiation at v=0.50 a.u., a difference that is mirrored in the stopping power (Figure 2), and all holes are formed within 10 eV of HOMO. Interestingly, under carbon ion irradiation, a large percentage (~18%) of the generated holes were on the phosphate side chain (see S17 in Supporting Information) at v=0.50 a.u. for the Base path, while neither proton nor $\alpha$-particle irradiation showed significant contributions from the phosphate side chains (less than 5%).

### 4. Conclusion:

In ion beam therapy, high-energy ions are used as ionizing radiation. Electronic excitations are induced as irradiating ions penetrate and transfer their kinetic energy into the target material. Our first-principles simulation of this electronic stopping process for DNA in water has shown that high-energy $\alpha$-particles and carbon ions indeed yield a higher energy transfer rate, measured by the electronic stopping power, than protons as the irradiating ion. At the same time, the observed increase in the electronic stopping power for the $\alpha$-particles and carbon ions does not follow the linear response theory description. Even when the velocity dependence of the irradiating ion's charge is accounted for, the discrepancies of the linear response theory model cannot be reconciled.

The simulations also show that for all ions the electronic excitation is highly localized around the projectile/irradiating ion paths, confirming the excitation remains conformal even for carbon ions.[8, 73] Additionally, as was previously observed for the case of proton irradiation[39], significantly more energy is transferred onto the sugar-phosphate side chains than onto the nucleobases for both the $\alpha$-particles and carbon ions. At the same time, differences in the electronic stopping power between the radiation ion types could not be attributed simply to the number of the generated holes. The increased number of highly energetic holes, formed in deep energy states, also contributes significantly to the larger electronic stopping power under heavier $\alpha$-particle and carbon ion irradiation.

In the context of ion beam therapy, it is not the larger stopping power *per se* that one should aim for by assuming the number of generated holes is proportional to the stopping power. Our simulations show a much more complicated picture in terms of how these generated holes might act as the source of oxidative DNA damage. Energetics and spatial characteristics of these generated holes show that carbon ion irradiation indeed generates the largest number of highly energetic holes on the sugar-phosphate side chains but not in a simple trend one might expect from the stopping power.

Our first-principles theoretical work here has revealed key details of the DNA electronic excitation under $\alpha$-particle and carbon ion irradiation in water, yielding atomistic insights into key characteristics associated with heavier ions, particularly carbon ions.[11, 74-75] These key results contribute to building a comprehensive molecular-level understanding of DNA damage under ion irradiation.

## ASSOCIATED CONTENT

Supporting Information.

The Supporting Information is available free of charge on the ACS Publications website.

Supporting Information contains additional computational details, initial charge state determinations, electronic stopping power calculation details and comparisons, velocity dependence of Wannier center displacements and spreads, time dependence of DNA hole populations and energetics and low velocity comparisons. (PDF)

## AUTHOR INFORMATION

**Corresponding Author**
*ykanai@unc.edu

**ORCID**
Christopher Shepard: 0000-0002-3604-414X.

**Notes**
The authors declare no competing financial interest.

## ACKNOWLEDGMENT

The authors would like to thank Dr. Yi Yao and Dr. Dillon C. Yost for helpful discussions. The work is supported by National Science Foundation Grant No. CHE-1954894 (C. S. and Y. K.). The QB@LL code used in this work implements theoretical formalisms developed under NSF Grants No. OAC-2209858 and No. CHE-1954894. An award of computer time was provided by the Innovative and Novel Computational Impact on Theory and Experiment (INCITE) program. This research used resources of the Argonne Leadership Computing Facility, which is a DOE Office of Science User Facility supported under Contract No. DE- AC02-06CH11357.